\documentclass[prd,showpacs,showkeys,nofootinbib,floatfix, 
               fleqn,preprint,12pt,tightenlines]{revtex4-1} 

\usepackage{amsmath,amssymb,revsymb,graphicx,dcolumn}

\newcommand{\beq}{\begin{equation}}
\newcommand{\eeq}{\end{equation}}
\newcommand{\beqa}{\begin{eqnarray}}
\newcommand{\eeqa}{\end{eqnarray}}
\newcommand{\bsubeqs}{\begin{subequations}}
\newcommand{\esubeqs}{\end{subequations}}

\begin{document}

\begin{widetext}
\noindent JETP Lett. \textbf{109}, 364 (2019)  
\hfill    arXiv:1812.07046 
\newline\vspace*{-2mm}
\end{widetext}

\title{Tetrads and q-theory}
\vspace*{1mm}

\author{F.R. Klinkhamer}
\email{frans.klinkhamer@kit.edu}

\affiliation{Institute for Theoretical Physics,
Karlsruhe Institute of Technology (KIT),\\
76128 Karlsruhe, Germany}

\author{G.E. Volovik}
\email{volovik@ltl.tkk.fi}
\affiliation{\mbox{Low Temperature Laboratory, Department of Applied Physics,}\\
\mbox{Aalto University, PO Box 15100, FI-00076 Aalto, Finland,}\\
and\\
\mbox{Landau Institute for Theoretical Physics,
Russian Academy of Sciences,}\\
\mbox{Kosygina 2, 119334 Moscow, Russia}\vspace*{4mm}}

\begin{abstract}
\vspace*{1mm}\noindent
As the microscopic structure
of the deep relativistic quantum vacuum
is unknown, a phenomenological approach ($q$-theory)
has been proposed to describe the vacuum degrees of freedom 
and the dynamics of the vacuum energy after the Big Bang.  
The original $q$-theory was based on a four-form field strength from a
three-form gauge potential.
However, this realization of $q$-theory, just as others suggested so far, 
is rather artificial and does not take into account the fermionic nature 
of the vacuum. We now propose a more physical realization of the $q$-variable. 
In this approach, we assume that the vacuum has the properties of a
plastic (malleable) fermionic crystalline medium.  The new approach
unites general relativity and
fermionic microscopic (trans-Planckian) degrees of freedom,
as the approach involves both the tetrad of standard gravity 
and the elasticity tetrad of the hypothetical vacuum crystal. 
This approach also allows for the description of possible topological phases 
of the quantum vacuum.\vspace*{-0mm}
\end{abstract}




\maketitle


\section{Introduction}
\label{sec:Intro}

The $q$-theory framework~\cite{KlinkhamerVolovik2008a,KlinkhamerVolovik2016-brane}
provides a general phenomenological approach to the
dynamics of vacuum energy, which may be useful for the resolution of
problems related to the cosmological constant in the Einstein equation
(a brief review of $q$-theory appears in Appendix A of
Ref.~\cite{KlinkhamerVolovik2011-review}).
The advantage of $q$-theory is that, at the classical level,
the field equations
of the theory essentially do not depend on the detailed microscopic
(trans-Planckian) origin of the $q$-field.
In the classical limit, the field equations of $q$-theory
(i.e., the equation for the microscopic variable $q$
describing the quantum vacuum and the modified Einstein equation
for the metric) are universal.

The $q$-theory approach to the cosmological constant problem
aims to describe the decay of the vacuum energy density from an initial
Planck-scale value to the present value of the cosmological constant.
However, the correct
description of this decay requires the quantum version of $q$-theory,
which may be different for different classes of realizations
of the $q$-variable
(see Sec.~1 in Ref.~\cite{KlinkhamerSavelainenVolovik2017}
for a general discussion of quantum-dissipative effects and
Ref.~\cite{KlinkhamerVolovik2016-Lambda-cancellation} for a sample calculation).

Up till now, our discussions of $q$-theory have primarily
used the nonlinear theory of a four-form field strength
from a three-form gauge potential
(the linear theory of the vacuum energy in
terms of the four-form field strength has been considered
by Hawking~\cite{Hawking1984}, in particular).
However, the four-form field strength,
though useful for the construction
of the general phenomenological equations for the quantum vacuum,
is rather abstract. The physical origin of such a field is not clear.
Instead, we must look for another variable, which may have a
direct relation to the underlying microscopic physics of the quantum vacuum.

Here, we propose a physical realization of the vacuum as a plastic (malleable) medium, which locally has the structure of a $(3+1)$-dimensional fermionic crystal. In this realization, the corresponding vacuum variable $q$ is expressed in terms of both the gravitational tetrad and the elasticity tetrad of the underlying crystal. The new realization in terms of tetrad fields is more appropriate for the quantum theory of the fermionic vacuum of our Universe than the realization with a bosonic four-form field strength.

As distinct from the conventional gravitational tetrads,
the elasticity tetrads have dimensions of inverse length
or inverse time. This allows for the construction of topological
Chern-Simons terms describing the mixed gravity-elasticity anomaly and
the mixed gauge-elasticity anomaly of the fermionic vacuum.

Throughout, we use natural units with $c=\hbar=1$
and take the metric signature $(-+++)$.

\section{Gravity tetrad}
\label{sec:Gravity-tetrad}

The tetrad formalism of torsion-less gravity is given
by the following equations:
\bsubeqs\label{eq:MetricTetrad-eqs}
\begin{eqnarray}
 g_{\mu\nu} &=&\eta_{ab}\,e_\mu^a\, e_\nu^b \,,
\label{eq:MetricTetrad}
\\[2mm]
 \nabla_\mu\, g^{\mu\nu}&=&0\,,
\label{eq:Derivative-metric}
\\[2mm]
D_\mu\, e^a_\nu  &\equiv&
\nabla_\mu\, e^a_\nu + \omega^a_{\mu b}\,e^b_\nu=0 \,,
\label{eq:Derivative-tetrad}
\end{eqnarray}
\esubeqs
where $\nabla_\mu$ is the standard covariant derivative of general relativity
and $\omega^a_{\mu b}$ the spin connection,
\begin{equation}
\omega^a_{\mu b}=e^a_\nu\, \nabla_\mu\, e ^\nu_b \,.
\label{eq:spin-connection}
\end{equation}
Given the tetrad $e_\mu^a$,
the inverse tetrad $e^\mu_a$ is defined by
\bsubeqs\label{eq:InverseTetrad-def1-def2}
\begin{eqnarray}
e_\mu^a \, e^\mu_b &=&\delta_{a}^{b} \,,
\label{eq:InverseTetrad-def1}
\end{eqnarray}
or, equivalently,
\begin{eqnarray}
e_\mu^a \, e^\nu_a &=&\delta_{\mu}^{\nu} \,.
\label{eq:InverseTetrad-def2}
\end{eqnarray}
\esubeqs
An introduction to the tetrad formalism of gravitation
can be found in, e.g., Sec.~12.5 of Ref.~\cite{Weinberg1972}.

\section{Elasticity tetrad}
\label{sec:Elasticity-tetrad}

In the present article, we interpret the vacuum
as a plastic (malleable) fermionic crystalline medium.
The ``vacuum crystal'' could, in principle, also have
bosonic ``atoms,'' but, for definiteness, we focus on the
fermionic case.
At each point of spacetime, we have a local system of four deformed
crystallographic  manifolds of constant phase
$X^a(x)=2\pi n^a$, for $n^a \in \mathbb{Z}$
with $a=0,\,1,\,2,\,3$. (For the usual atomic crystals
in 3-space, the index $a$ runs over $1,\,2,\,3$.
See Sec.~6 in Ref.~\cite{DzyaloshinskiiVolovick1980} for
a general discussion of the elasticity theory of crystals
and Sec.~3.2 in Ref.~\cite{NissinenVolovik2018a} for a succinct summary
of the role of the phase functions $X^a$.)

In addition to the conventional tetrad $e_\mu^a$ of gravity,
we then introduce the following elasticity tetrad $E^{a}_\mu$
\mbox{(cf. Refs.~\cite{DzyaloshinskiiVolovick1980,NissinenVolovik2018a,%
NissinenVolovik2018b})}:
\begin{equation}
E^{~a}_\mu(x)= D_\mu\, X^a(x)\,,
\label{reciprocal}
\end{equation}
where both indices $a$ and $\mu$ take values
from the set $\{0,\,1,\,2,\,3\}$.
Invariance under the local $SO(1,\, 3)$ group of rotations
is implemented by defining
\begin{equation}
 D_\mu\, X^a  \equiv \nabla_\mu\, X^a+ \omega^a_{\mu b}\,X^b
 = \partial_\mu\, X^a+ \omega^a_{\mu b\,}X^b \,.
\label{eq:VectorDerivative}
\end{equation}
Note that the dimensionalities of the elasticity
tetrads $E^{~a}_i$ and $E^{~a}_0$ are
inverse length and inverse time, respectively.

\section{Q-field realization by tetrads}
\label{sec:q-field realization by tetrads}

Let us, now,
consider the $q$-theory description
of the quantum vacuum~\cite{KlinkhamerVolovik2008a}
by assuming that the vacuum energy density $\epsilon(q)$ in the action
depends on the following type of $q$-field:
\begin{equation}
q(x) =\frac{1}{4}\, e^\mu_a(x) \, E_\mu^a(x)  \,,
\label{eq:q-field}
\end{equation}
in terms of the inverse of the gravity tetrad $e_\mu^a$
from \eqref{eq:MetricTetrad-eqs}
and the elasticity tetrad $E_\mu^a$  from \eqref{reciprocal}.
The action in its simplest form is then given by
\begin{equation}
S=
\int_{\mathbb{R}^4}
\,d^4x\, e\,
\left(\frac{R}{16\pi G_N} +\epsilon(q)
\right) \,,
\label{eq:action-q}
\end{equation}
where $R$ is the Ricci curvature scalar and $e$ the tetrad determinant,
\begin{equation}
 e\equiv \det e^a_\mu \,.
\label{TetradDeterminant}
\end{equation}
The pure-gravity part of the action \eqref{eq:action-q}
contains the integral
\begin{equation}
\int_{\mathbb{R}^4}\,d^4x\, e\,R =
\int_{\mathbb{R}^4}\,d^4x\, e\,e^\mu_a e^\nu_b\, F^{ab}_{\mu\nu} \,,
\label{eq:R}
\end{equation}
with a curvature two-form given by
\begin{eqnarray}
F^{ab} = d\,\omega^{ab} + \omega_c^a \wedge\omega^{bc} \,,
\label{eq:F}
\end{eqnarray}
in terms of the spin connection \eqref{eq:spin-connection}.

Variation of the action \eqref{eq:action-q} over $e_a^\mu$
gives the Einstein equation~\cite{Weinberg1972},
\begin{eqnarray}
\label{eq:field-eqs-Einstein}
R_{\mu\nu} - \frac12\, g_{\mu\nu}\,R
 &=&  8\pi G_{N} \, \rho_{V}(q)  \, g_{\mu\nu} \,,
\end{eqnarray}
where $\rho_{V}(q)$ will be discussed shortly,
and variation over $X^a$ gives
the following differential equation for $q$
(which is both a coordinate scalar and a Lorentz scalar):
\begin{eqnarray}
\label{eq:nabla-depsilondq}
\partial_\mu \left(\frac{d\epsilon(q)}{dq}\right) &=& 0 \,,
\end{eqnarray}
where \eqref{eq:Derivative-tetrad} has been used.
The vacuum energy density $\rho_{V}(q)$,
which enters the Einstein equation \eqref{eq:field-eqs-Einstein}
through a cosmological-constant-type term, is given by
\begin{equation}
\rho_V(q) \equiv \epsilon(q) - q \,\frac{d\epsilon(q)}{dq} \,,
\label{eq:rhoV}
\end{equation}
with an extra term $- q\,d\epsilon/dq$
tracing back to the $e^\mu_a$ dependence of
the $q$-realization \eqref{eq:q-field}.

The field equation \eqref{eq:nabla-depsilondq} for $q$ has
the following general solution:
\begin{equation}
 \frac{d\epsilon(q)}{dq} = \mu ={\rm constant}\,,
\label{eq:mu}
\end{equation}
where the arbitrary constant $\mu$
(interpreted as a ``chemical potential''
in Ref.~\cite{KlinkhamerVolovik2008a})
is not to be confused with the spacetime index $\mu$.
The solution \eqref{eq:mu} allows us to rewrite the
gravitating vacuum energy density \eqref{eq:rhoV}
as $\rho_V(q) = \epsilon(q) - \mu\,q$.

The quantum vacuum in perfect equilibrium has a constant
nonzero value of the $q$-field,
\begin{equation}
q(x)=q_{0} ={\rm constant}\,,
\label{eq:q0}
\end{equation}
which gives a particular value $\mu_0$
for the constant $\mu$ in \eqref{eq:mu},
\begin{equation}
\mu_0 =
\left[\frac{d\epsilon(q)}{dq}\right]_{q=q_{0}} \,.
\label{eq:mu0}
\end{equation}
In addition, there are the following equilibrium conditions:
\bsubeqs\label{eq:equilibrium-conditions-a-b-c}
\beqa\label{eq:equilibrium-conditions-a}
\rho_{V}(q_{0}) &=& 0\,,
\\[2mm]
\label{eq:equilibrium-conditions-b}
\left[\frac{d\,\rho_{V}(q)}{d q}\right]_{q=q_{0}} &=& 0\,,  
\\[2mm]
\label{eq:equilibrium-conditions-c}
\left[\frac{d^2\,\rho_{V}(q)}{dq^2}\right]_{q=q_{0}} &>& 0\,. 
\eeqa
\esubeqs
These conditions have been discussed for $q$-theory in general
(see, e.g., Sec.~II B in Ref.~\cite{KlinkhamerVolovik2008a}).
First, recall that the
conditions \eqref{eq:equilibrium-conditions-a}
and \eqref{eq:equilibrium-conditions-b} result from the
self-adjustment of the conserved vacuum variable $q$
[having, as mentioned before, the
chemical potential $\mu=d\epsilon/dq$ and
the thermodynamically active vacuum energy density
$\rho_V(q) = \epsilon(q) - \mu\,q$],
as follows from the Gibbs--Duhem relation for an isolated
self-sustained system without external pressure.
Second, recall that condition \eqref{eq:equilibrium-conditions-c}
can be interpreted as having a positive isothermal compressibility.
A further important consequence of the
equilibrium conditions \eqref{eq:equilibrium-conditions-a-b-c}
will be discussed in the next section.

To summarize, we have obtained with \eqref{eq:q-field}
one further realization of the $q$-variable, in addition to the
four-form realization~\cite{KlinkhamerVolovik2008a}
and the brane realization~\cite{KlinkhamerVolovik2016-brane}.
The advantage of this new realization is that it has a more direct
physical origin.

\section{Discussion}
\label{sec:Discussion}

The action \eqref{eq:action-q} is special in the sense that
the only dependence on the ``vacuum crystal'' is via the
$q$-variable as defined by \eqref{eq:q-field}.
Another approach is to assume that the action depends on
the elasticity tetrad $E^{~a}_\mu(x)$
in a general way, but still obeys all the gauge symmetries.
One possible action density term would be, for example,
\beq\label{eq:action-density-term}
k\,\left[ \nabla^{\kappa} (X^a e^\lambda_a) \right]\,
\left[ \nabla^{\mu} (X^b e^\nu_b) \right] \,
g_{\kappa\mu}\,g_{\lambda\nu}\,,
\eeq
with a constant $k$.
Then, $q$ only appears via the equilibrium solution of
the field equation obtained by variation over $X^a$,
\begin{equation}
\Big[ D_\mu X^a\,\Big]_\text{equil.\;sol.}= q\, e_\mu^a \,,
\label{eq:equilibrium}
\end{equation}
for constant $q$.

In this approach with a general dependence of the
action on the elasticity tetrad, we expect that
the local Newtonian gravitational dynamics is ruined.
The reason is that we have from \eqref{eq:equilibrium} that
\begin{equation}
\big[q\,\big]_\text{equil.\;sol.}
= \frac{1}{4}\, e^\mu_a \,D_\mu\, X^a
= D_\mu \left(\frac{1}{4}\, e^\mu_a \, X^a \right)
\equiv  D_\mu\, \mathcal{A}^\mu
=  \nabla_\mu\, \mathcal{A}^\mu \,,
\label{eq:q-field2}
\end{equation}
where the composite vector field $\mathcal{A}^\mu(x)$
plays a similar role as 
the fundamental vector field $A^\mu(x)$ in Dolgov's 
discussion of the
cosmological constant problem~\cite{Dolgov1985,Dolgov1997}.
In fact, the
action density term \eqref{eq:action-density-term},
rewritten in terms of
$\mathcal{A}^\mu(x)$ from \eqref{eq:q-field2},
corresponds precisely to Dolgov's
term $[\nabla_{\mu} A_{\nu}]\,[\nabla^{\mu} A^{\nu}]$.

From \eqref{eq:q-field2},
we now observe that, in a cosmological context, a constant $q$
requires \mbox{$\mathcal{A}_0 \propto t$}.
But precisely the behavior $\mathcal{A}_0 \propto t$ is
the source of trouble for Newtonian gravity as
shown by Rubakov and Tinyakov~\cite{RubakovTinyakov2000}.
(It is possible to maintain
Newtonian gravity by the introduction of further $q$-type
fields, corresponding to different interpenetrating
vacuum crystals, but the theory looks rather
artificial~\cite{EmelyanovKlinkhamer2012a,EmelyanovKlinkhamer2012c,%
SantillanScornavacche2017}.)

This disaster of ruining  Newtonian gravity
is not expected to occur in the approach of
Sec.~\ref{sec:q-field realization by tetrads}.
The reason is two-fold. First, we observe that
the corresponding field equations do not explicitly carry
the composite vector field $\mathcal{A}^\mu(x)$
(as happens for the Dolgov theory;
cf. (A1) with $\zeta=1$ in Ref.~\cite{EmelyanovKlinkhamer2012a})
but only contain the fields $g_{\mu\nu}(x)$ and $q(x)$,
as shown by \eqref{eq:field-eqs-Einstein}, \eqref{eq:nabla-depsilondq},
and \eqref{eq:rhoV}.
Second, we observe that the Einstein equation
\eqref{eq:field-eqs-Einstein}
with $\rho_V \sim (q-q_0)^2$,
according to the equilibrium conditions \eqref{eq:equilibrium-conditions-a-b-c},
gives rise to the standard linearized Einstein equation
for a small perturbation around flat Minkowski
spacetime with $\rho_V(q_0)=0$.
Regarding the last observation, we refer
to Ref.~\cite{Klinkhamer2016-brief-report}
for a discussion of the special case of four-form $q$-theory
and to Sec.~2.2 in Ref.~\cite{Klinkhamer2016}
for a general discussion.

In closing, we remark that our setup with vacuum elasticity
resembles somewhat the setup of gravity as a
gauge-theory-squared~\cite{Anastasiou-etal2018}, where the metric is
considered to be the product of two Yang--Mills fields.
The elasticity tetrad plays a similar role as the Yang--Mills fields.
But, for us,
the elasticity tetrad describes the vacuum as a (3+1)-dimensional
fermionic crystalline medium, where dislocations correspond to torsion
and disclinations to curvature~\cite{DzyaloshinskiiVolovick1980}.
Such a medium may also have different nontrivial topological phases,
characterized by particular Chern--Simons-like terms
with prefactors determined by momentum-space topological
invariants~\cite{NissinenVolovik2018b}.

Another interpretation of the vacuum crystal as
described by an elasticity tetrad is that this crystal may
provide a dynamic realization of the prior metric which
enters a particular formulation of $q$-theory, where
the $q$-variable essentially equals the determinant
of the metric relative to the determinant
of a fixed prior metric~\cite{Klinkhamer2016}.

A further incentive for choosing the elasticity tetrad as a realization of the $q$-variable is based on the following observation. The quantum version of $q$-theory is sensitive to the particular realization of the $q$-field.  Assuming $q$-theory to be relevant, the comparison with experiment may then provide information on the detailed structure of the fermionic quantum vacuum and, in particular, on the types of quantum anomalies. The fermionic crystalline model of the vacuum is one of the possible structures of the deep fermionic vacuum, distinct from a structure described by  the abstract four-form field strength. This new fermionic structure gives, for example, rise to new types of quantum anomalies, where elasticity tetrads are mixed   with  gauge and spin-connection fields~\cite{NissinenVolovik2018b}.

\begin{acknowledgments}
The work of GEV has been supported by the European Research Council
(ERC) under the European Union's Horizon 2020 research and innovation
programme (Grant Agreement No. 694248).
\end{acknowledgments}


\end{document}